\documentclass[12pt,a4paper]{amsart}
\usepackage{amsmath,amssymb,a4wide,amsthm,ifthen,hyperref,xcolor,enumerate,enumitem}
\usepackage[utf8]{inputenc}

\usepackage[foot]{amsaddr}
\usepackage{graphicx}
\usepackage{url}
\usepackage{ulem}
\usepackage{natbib}
\usepackage{flafter} 
\usepackage{multirow}
\usepackage{float}
\usepackage{tabularx,ragged2e}
\usepackage{booktabs,ulem}
\usepackage{makecell}
\newcolumntype{Y}{>{\centering\arraybackslash}X}

\theoremstyle{remark}

\newcommand{\by}{\mathbf{y}}
\newcommand{\bX}{\mathbf{X}}
\newcommand{\bSigma}{\boldsymbol{\Sigma}}

\newcommand{\bU}{\mathbf{U}}
\newcommand{\bD}{\mathbf{D}}
\newcommand{\bbeta}{\boldsymbol{\beta}}

\newcommand{\bepsilon}{\boldsymbol{\epsilon}}

\newcommand{\bgamma}{\boldsymbol{\gamma}}
\newcommand{\bzero}{\mathbf{0}}
\newcommand{\bone}{\mathbf{1}}
\newcommand{\norm}[1]{\left\lVert#1\right\rVert}

\def\argmin{\mathop{\rm Argmin}\nolimits}

\begin{document}

\title[Identification of prognostic and predictive biomarkers]
{Identification of prognostic and predictive biomarkers in high-dimensional data with PPLasso}

\date{}

\author{Wencan Zhu}
\address{UMR MIA-Paris, AgroParisTech, INRAE, Universit\'e Paris-Saclay, 75005, Paris, France}
\email{wencan.zhu@agroparistech.fr}
\author{Céline Lévy-Leduc}
\address{UMR MIA-Paris, AgroParisTech, INRAE, Universit\'e Paris-Saclay, 75005, Paris, France}
\email{celine.levy-leduc@agroparistech.fr}
\author{Nils Ternès}
\address{Biostatistics and Programming department, Sanofi R\&D, 91380 Chilly Mazarin, France}
\email{nils.ternes@sanofi.com}

\keywords{variable selection; highly correlated predictors; genomic data}

\maketitle

\begin{abstract}
In clinical development, identification of prognostic and predictive biomarkers is essential to precision medicine. Prognostic biomarkers can be useful for anticipating
the prognosis of individual patients, and predictive biomarkers can be used to identify patients more likely to benefit from a given treatment. Previous researches were mainly focused on clinical characteristics, and the use of genomic data in such an area is hardly studied. A new method is required to simultaneously select prognostic and predictive biomarkers in high dimensional genomic data where biomarkers are highly correlated. We propose a novel approach called PPLasso (Prognostic Predictive Lasso) integrating prognostic and predictive effects into one statistical model. PPLasso also takes into account the correlations between biomarkers that can alter the biomarker selection accuracy. Our method consists in transforming the design matrix to remove the correlations between the biomarkers before applying the generalized Lasso.
In a comprehensive numerical evaluation, we show that PPLasso outperforms the Lasso type approaches on both prognostic and
predictive biomarker identification in various scenarios. Finally, our method is applied to publicly available
transcriptomic data from clinical trial RV144. Our method is implemented in the \texttt{PPLasso} R package which is available from the Comprehensive
R Archive Network (CRAN).
\end{abstract}

\section{Introduction}

With the advancement of precision medicine, there has been an increasing interest in identifying prognostic or predictive biomarkers in clinical development. A prognostic biomarker informs about a likely clinical outcome (e.g., disease recurrence, disease progression, death) in the absence of therapy or with a standard therapy that patients are likely to receive, while a predictive biomarker is associated with a response or a lack of response to a specific therapy. \cite{Ballman2015} and  \cite{Clark2008} provided a comprehensive explanation and concrete examples to distinguish  prognostic from predictive biomarkers, respectively.

Concerning the biomarker selection, the high dimensionality of genomic data is one of the main challenges as explained in \cite{Fan2006}. To identify effective biomarkers in high-dimensional settings, several approaches can be considered including hypothesis-based tests described in \cite{Ttest}, wrapper approaches \textcolor{black}{proposed in \cite{Saeys2007}}, and penalized approaches such as Lasso designed by \cite{Lasso} among others. Hypothesis-based tests consider each biomarker independently
and thus ignore potential correlations between them. Wrapper approaches often show high risk of overfitting and are computationally expensive for high-dimensional data as explained in \cite{Stepwise}. More efforts have been devoted to penalized methods given their ability to automatically perform variable selection and coefficient estimation simultaneously as highlighted in \cite{Fan2009}. However, Lasso showed some potential drawbacks when biomarkers are highly correlated. Particularly, when the Irrepresentable Condition (IC) \textcolor{black}{proposed by \cite{Zhao:2006}} is violated, Lasso can not guarantee to correctly identify true effective biomarkers. In genomic data, biomarkers are usually highly correlated such that this condition can hardly be satisfied, see \cite{Wang:2018}. Several methods have been proposed to adress this issue. \textcolor{black}{Elastic Net \citep{Zou2005} combines the $\ell_1$ and $\ell_2$ penalties and is particularly effective in tackling correlation issues and can generally outperform Lasso. Adaptive Lasso \citep{Zou2006} proposes to assign adaptive weights for penalizing different coefficients in the $\ell_1$ penalty, and its oracle property was demonstrated.} \cite{holp:2016} proposed the HOLP approach which consists in removing the correlation between the columns of the design matrix; \cite{Wang:2018} proposed to handle the correlation by assigning similar weights to correlated variables \textcolor{black}{in their approach called Precision Lasso}; \cite{Zhu2021} proposed to remove the correlations by applying a whitening transformation to the data before using the \textcolor{black}{generalized Lasso criterion designed by \cite{tibshirani:2011}.} 

The challenge of finding prognostic biomarkers has been extensively explored with previously introduced methods, however, the discovery of predictive biomarkers has seen much less attention.  Limited to binary endpoint, \cite{Foster2011} proposed to first predict response probabilities for treatment and use this probability as the response in a classification problem to find effective biomarkers. \cite{Tian2012} proposed a new method to detect interaction between the treatment and the biomarkers by modifying the covariates. This method can be implemented on continuous/binary/time-to-event endpoint. \cite{SIDES2011} proposed a method called SIDES, which adopts a recursive partitioning algorithm for screening treatment-by-biomarker interactions. This method was further improved  in \cite{SIDES2014} by adding another step of preselection on predictive biomarkers based on variable importance. The method was demonstrated with continuous endpoint. More recently, \cite{Sechidis2018} applied  approaches coming from information theory for ranking biomarkers on their prognostic/predictive strength. Their method is applicable only for binary or time-to-event endpoint. Moreover, all of these methods were assessed under the situation where the sample size is relatively large and the number of biomarkers is limited, which is hardly the case for genomic data.

In the literature mentioned above, the authors focused on one of the problematic of identifying prognostic or predictive biomarkers, but rarely on both. Even if predictive biomarkers is of major importance for identifying patients more likely to benefit from a treatment, the identification of prognostic biomarkers is also key in this context. Indeed, the clinical impact of a treatment can be judged only with the knowledge of the prognosis of a patient. It is thus of importance to reliably predict the prognosis of patients to assist treatment counseling \citep{Windeler2000}.
In this paper, we propose a novel approach called PPLasso (Prognostic Predictive Lasso) to simultaneously identify prognostic and predictive biomarkers in a high dimensional setting with continuous endpoints, as presented in Section \ref{sec:methods}. 
Extensive numerical experiments are given in Section \ref{sec:numexp} to assess
the performance of our approach and to compare it to other methods. PPLasso is also applied to the clinical trial RV144
in Section \ref{sec:real}. Finally, we give concluding remarks in Section \ref{sec:conclusion}.

\section{Methods}\label{sec:methods}

In this section, we propose a novel approach called PPLasso (Predictive Prognostic Lasso) which consists in writing the identification of predictive and prognostic biomarkers as a variable selection problem in an ANCOVA (Analysis of Covariance) type model mentioned for instance in \cite{faraway:2002}.

\subsection{Statistical modeling}

Let $\by$ be a continuous response or endpoint and $t_1$, $t_2$ two treatments.
Let also $\bX_{1}$ (resp. $\bX_{2}$) denote the design matrix for the $n_1$ (resp. $n_2$) patients with treatment $t_1$ (resp. $t_{2}$),
each containing measurements on $p$ candidate biomarkers:
{\small{
\begin{equation}\label{eq:X1_X2}
\bX_1= \begin{bmatrix}
   X_{11}^{1} & X_{11}^{2} & \ldots & X_{11}^{p} \\
   X_{12}^{1} & X_{12}^{2} & \ldots & X_{12}^{p} \\
   ...\\
   X_{1n_{1}}^{1} & X_{1n_{1}}^{2} & \ldots & X_{1n_{1}}^{p} 
\end{bmatrix},\bX_2=\begin{bmatrix}
   X_{21}^{1} & X_{21}^{2} & \ldots & X_{21}^{p} \\
   X_{22}^{1} & X_{22}^{2} & \ldots & X_{22}^{p} \\
   ...\\
   X_{2n_{2}}^{1} & X_{2n_{2}}^{2} & \ldots & X_{2n_{2}}^{p}
   \end{bmatrix}.
 \end{equation}
 }}
To take into account the potential correlation that may exist between the biomarkers in the different treatments, we shall assume that the rows of $\bX_{1}$
(resp. $\bX_{2}$) are independent centered Gaussian random vectors with a covariance matrice equal to $\boldsymbol{\Sigma_1}$ (resp. $\boldsymbol{\Sigma_2}$).

To model the link that exists between $\by$ and the different
types of biomarkers we propose using the following model:
\begin{equation}\label{eq:y}
 \by=\begin{pmatrix} y_{11} \\ y_{12}\\\vdots\\ y_{1n_{1}}\\
 y_{21} \\ y_{22}\\\vdots\\ y_{2n_{2}}
 \end{pmatrix}
 =\bX
    \begin{pmatrix} \alpha_1 \\ \alpha_2 \\ \beta_{11} \\ \beta_{12}\\ \vdots\\\beta_{1p} \\
    \beta_{21} \\ \beta_{22}\\\vdots\\\beta_{2p}
 \end{pmatrix} +
 \begin{pmatrix} \epsilon_{11} \\ \epsilon_{12}\\ \vdots \\\epsilon_{1n_{1}}\\
\epsilon_{21} \\ \epsilon_{22}\\ \vdots \\\epsilon_{2n_{2}}
 \end{pmatrix},
\end{equation}
where $(y_{i1},\dots,y_{in_i})$ corresponds to the response of patients with treatment $t_i$, $i$ being equal to 1 or 2,
 $$
 \bX=\begin{bmatrix}
    1 & 0 & X_{11}^{1} & X_{11}^{2} & \ldots & X_{11}^{p} &   0 & 0 & \ldots & 0  \\
    1 & 0 &  X_{12}^{1} & X_{12}^{2} & \ldots & X_{12}^{p} & 0 & 0 & \ldots & 0  \\
  \vdots &  \vdots & \vdots& \vdots& &  \vdots & & & \\
  1 &  0 & X_{1n_{1}}^{1} & X_{1n_{1}}^{2} & \ldots & X_{1n_{1}}^{p} & 0 & 0 & \ldots & 0  \\
    0 &  1 & 0 & 0 & \ldots & 0 &  X_{21}^{1} & X_{21}^{2} & \ldots & X_{21}^{p} \\
    0 &  1 &  0 & 0 & \ldots & 0 &    X_{22}^{1} & X_{22}^{2} & \ldots & X_{22}^{p} \\
     \vdots &  \vdots & \vdots& \vdots& & \vdots& \vdots&\vdots & &\vdots\\
    0 & 1 & 0 & 0 & \ldots & 0 & X_{2n_{2}}^{1} & X_{2n_{2}}^{2} & \ldots & X_{2n_{2}}^{p}
   \end{bmatrix},
   $$
with $\alpha_1$ (resp. $\alpha_2$) corresponding to the effects of treatment $t_1$ (resp. $t_2$). 
 Moreover, $\bbeta_1=(\beta_{11}, \beta_{12}, \ldots, \beta_{1p})'$ (resp. $\bbeta_2=(\beta_{21}, \beta_{22}, \ldots, \beta_{2p})'$) are the coefficients associated to each of the $p$ biomarkers in treatment $t_1$ (resp. $t_2$) group, $'$ denoting the matrix transposition
 and $\epsilon_{11},\dots,\epsilon_{2n_2}$ are standard independent Gaussian random variables independent of $\bX_{1}$ and $\bX_{2}$. When $t_1$ stands for the standard treatment or placebo, prognostic biomarkers are defined as those having non-zero coefficients in $\bbeta_{1}$. According to the definition of prognostic biomarkers, their effect should indeed be demonstrated in the absence of therapy or with a standard therapy that patients are likely to receive. On the other hand, predictive biomarkers are defined as those having non-zero coefficients in $\bbeta_{2}-\bbeta_{1}$ because they aim to highlight different effects between two different treatments.\\
 
%
Model (\ref{eq:y}) can be written as:
\begin{equation}\label{lm_simp}
    \by=\bX\bgamma+\bepsilon,
\end{equation}
with $\bgamma=(\alpha_1,\alpha_2, \bbeta_{1}', \bbeta_{2}')'$.
The Lasso penalty is a well-known approach to estimate coefficients with a sparsity enforcing constraint allowing variable selection by estimating
some coefficients by zero. It consists in minimizing the following
penalized least-squares criterion (\cite{Lasso}):
\begin{equation}
\label{eq:LASSO}
 \frac{1}{2}\norm{\by-\bX\bgamma}_{2}^{2}+\lambda\norm{\bgamma}_{1}, 
\end{equation}
where $\norm{\textbf{u}}_{2}^2=\sum_{i=1}^n u_i^2$ and $\norm{\textbf{u}}_{1}=\sum_{i=1}^n |u_i|$ for $\textbf{u}=(u_1,\dots,u_n)$. A different sparsity constraint
was applied to $\bbeta_1$ and $\bbeta_{2}-\bbeta_{1}$ to allow different sparsity levels.
Hence we propose to replace the penalty $\lambda\norm{\bgamma}_{1}$ in (\ref{eq:LASSO}) by 
\begin{equation}
    \lambda_{1}\norm{\bbeta_{1}}_{1}+\lambda_{2}\norm{\bbeta_{2}-\bbeta_{1}}_{1}.
\end{equation}
%
Thus, a first estimator of $\bgamma$ could be found by minimizing the following criterion with respect to $\bgamma$:
\begin{equation}
\label{eq:newW}
\frac{1}{2}\norm{\by-\bX\bgamma}_{2}^{2}+\lambda_{1}\norm{\begin{bmatrix}
\bzero_{p,1} &  \bzero_{p,1} & D_{1}\\\bzero_{p,1} & \bzero_{p,1} & \frac{\lambda_{2}}{\lambda_{1}}D_{2}\end{bmatrix}\bgamma}_{1},
\end{equation}
\textcolor{black}{where $D_1=[\mathbf{\textrm{Id}}_{p}, \bzero_{p,p}]$ and $D_2=[-\mathbf{\textrm{Id}}_{p},\mathbf{\textrm{Id}}_{p}]$, with $\mathbf{\textrm{Id}}_{p}$ denoting the identity matrix of size $p$ and $\bzero_{i,j}$ denoting a matrix having $i$ rows and $j$ columns
  and containing only zeros.}
%
However, since the inconsistency of Lasso biomarker selection is originated from the correlations between the biomarkers, we propose to remove the correlation by
``whitening'' the matrix $\bX$. More precisely, we consider $\widetilde{\bX}=\bX\bSigma^{-1/2}$, where
\begin{equation}
\label{eq:Sigma}
\bSigma=
\begin{bmatrix}
 1 & 0 &  0 & 0\\
 0 & 1 & 0 & 0 \\
 
 0 & 0 & \bSigma_{1} & 0 \\
 0 & 0 & 0&  \bSigma_{2}
\end{bmatrix}
\end{equation} 
and define $\bSigma^{-1/2}$ by replacing in (\ref{eq:Sigma}) $\bSigma_{i}$ by $\bSigma_{i}^{-1/2}$, where $\bSigma_{i}^{-1/2}=\bU_{i}\bD_{i}^{-1/2}\bU_{i}^{T}$,
$\bU_{i}$ and $\bD_{i}$ being the matrices involved in the spectral decomposition of $\bSigma_{i}$ for $i=1$ or 2. With such a transformation the columns of $\widetilde{\bX}$
are decorrelated and Model (\ref{lm_simp}) can be rewritten as follows:
\begin{equation}\label{lm_w}
    \by= \widetilde{\bX}\widetilde{\bgamma}+\bepsilon
\end{equation} 
where $\widetilde{\bgamma}=\bSigma^{1/2}\bgamma$. The objective function (\ref{eq:newW}) thus becomes:
\begin{equation}
\label{eq:newW_trans}
L_{\lambda_1,\lambda_2}^{\textrm{PPLasso}}(\widetilde{\bgamma})=\frac{1}{2}\norm{\by-\widetilde{\bX}\widetilde{\bgamma}}_{2}^{2}+\lambda_{1}\norm{\begin{bmatrix}
\bzero_{p,1} &  \bzero_{p,1} & D_{1}\\\bzero_{p,1} & \bzero_{p,1} & \frac{\lambda_{2}}{\lambda_{1}}D_{2}\end{bmatrix}\bSigma^{-1/2}\widetilde{\bgamma}}_{1}.
\end{equation}

\subsection{Estimation of $\widetilde{\bgamma}$}\label{subsec:gamma_tilde}
\textcolor{black}
{Let us define a first estimator of $\widetilde{\bgamma}=(\widetilde{\alpha}_1,\widetilde{\alpha}_2,\widetilde{\bbeta}_1',\widetilde{\bbeta}_2')$ as follows:
\begin{equation}\label{eq:gamma0}
  \widehat{\widetilde{\bgamma}}_{0}(\lambda_1,\lambda_2)=(\widehat{\widetilde{\alpha}}_{1},\widehat{\widetilde{\alpha}}_{2}, \widehat{\widetilde{\bbeta}}_{10}',
  \widehat{\widetilde{\bbeta}}_{20}')=\argmin_{\widetilde{\bgamma}}  L_{\lambda_1,\lambda_2}^{\textrm{PPLasso}}(\widetilde{\bgamma}),
\end{equation}
for each fixed $\lambda_1$ and $\lambda_2$.
To better estimate $\widetilde{\bbeta}_1$ and $\widetilde{\bbeta}_2$, a thresholding was applied to
$\widehat{\widetilde{\bbeta}}_{0}(\lambda_1,\lambda_2)=(\widehat{\widetilde{\bbeta}}_{10}(\lambda_1,\lambda_2)',\widehat{\widetilde{\bbeta}}_{20}(\lambda_1,\lambda_2)')'$.
For $K_1$ (resp. $K_2$) in $\{1,\ldots,p\}$, let $\textrm{Top}_{K_1}$ (resp. $\textrm{Top}_{K_2}$) be the set of indices corresponding to the $K_1$ (resp. $K_2$)
largest values of the components of $|\widehat{\widetilde{\bbeta}}_{10}(\lambda_1,\lambda_2)|$ (resp. $|\widehat{\widetilde{\bbeta}}_{20}(\lambda_1,\lambda_2)|$),
then the estimator of $\widetilde{\bbeta}=(\widetilde{\bbeta}_1',\widetilde{\bbeta}_2')$ after the correction is denoted by
$\widehat{\widetilde{\bbeta}}(\lambda_1,\lambda_2)=(\widehat{\widetilde{\bbeta}}_{1}^{(\widehat{K}_1)}(\lambda_1,\lambda_2),
\widehat{\widetilde{\bbeta}}_{2}^{(\widehat{K}_2)}(\lambda_1,\lambda_2))$
where the $j$th component of $\widehat{\widetilde{\bbeta}}_{i}^{(K_i)}(\lambda_1,\lambda_2)$, for $i=1$ or 2, is defined by:
\begin{equation}\label{eq:beta_tilde_thresh}
	\widehat{\widetilde{\bbeta}}_{ij}^{(K_i)}(\lambda_{1}, \lambda_{2})=
	\begin{cases}
	  \widehat{\widetilde{\bbeta}}_{i0j}(\lambda_{1}, \lambda_{2}), & j \in \textrm{Top}_{K_i} \\
	 \textrm{$K_1$th largest value of } |\widehat{\widetilde{\bbeta}}_{i0j}(\lambda_{1}, \lambda_{2})| , & j \not\in \textrm{Top}_{K_i}.
	\end{cases}
      \end{equation}
      }
      Note that the corrections are only performed on $\widehat{\widetilde{\bbeta}}_0$, the estimators $\widehat{\widetilde{\alpha}}_{1}$ and
      $\widehat{\widetilde{\alpha}}_{2}$ were not modified.  The choice of $K_1$ and $K_2$ will be
      explained in Section \ref{K1K2M}.

\begin{figure}[!h]
\centering
\includegraphics[scale=0.5]{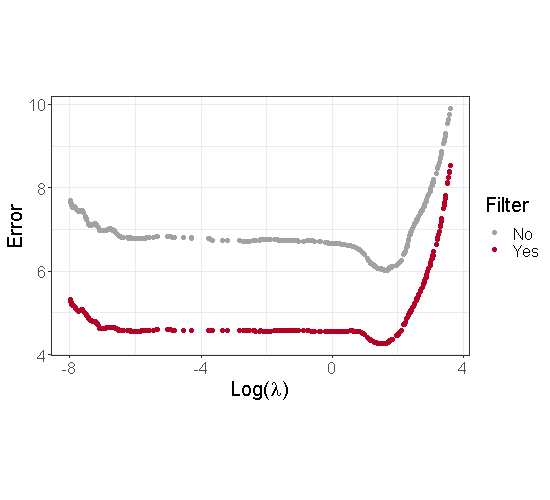}
\vspace{-18mm}
 \caption{\textcolor{black}{Estimation error $\norm{\widehat{\widetilde{\bbeta}}_{0}(\lambda)-\widetilde{\bbeta}}_{2}$ (gray) and
  $\norm{\widehat{\widetilde{\bbeta}}(\lambda)-\widetilde{\bbeta}}_{2}$ (red) for all $\lambda$.}
  \label{fig:gamma_tilde}}
\end{figure}

      To illustrate the interest of using a thresholding step, we generated a dataset based on Model \ref{lm_simp} with parameters described in Section
      \ref{setting} and $p=500$. Moreover, to simplify the graphical illustrations, we focus on the case where $\lambda_{1}=\lambda_{2}=\lambda$. 
      Figure \ref{fig:gamma_tilde} displays the estimation error associated to the estimators of $\widetilde{\bbeta}(\lambda)$ before and after the thresholding. We can see from this figure that the estimation of $\widetilde{\bbeta}(\lambda)$ is less biased after the correction.
      \textcolor{black}{Moreover, we observed that this thresholding strongly improves the final estimation of $\bgamma$ and the variable selection performance
        of our method.}

\subsection{Estimation of $\bgamma$}

With $\widehat{\widetilde{\bbeta}} = (\widehat{\widetilde{\bbeta}}_{1}', \widehat{\widetilde{\bbeta}}_{2}')$,
  the estimators of $\bbeta_1$ and $\bbeta_2-\bbeta_1$ can be obtained by
  $\widehat{\bbeta}_{10}=\boldsymbol{\Sigma}_{1}^{-1/2}\widehat{\widetilde{\bbeta}}_{1}$ and
  $(\widehat{\bbeta}_{20}-\widehat{\bbeta}_{10})=\boldsymbol{\Sigma}_{2}^{-1/2}\widehat{\widetilde{\bbeta}}_{2}-\boldsymbol{\Sigma}_{1}^{-1/2}\widehat{\widetilde{\bbeta}}_{1}$. As previously, another thresholding was applied to $\widehat{\bbeta}_{10}$ and $\widehat{\bbeta}_{20}$: for $i=1$ or 2,
%
%
\begin{equation}\label{eq:beta_hat1}
	\widehat{\bbeta}_{ij}^{(M_i)}(\lambda_{1}, \lambda_{2})=
	\begin{cases}
	  \widehat{\bbeta}_{i0j}(\lambda_{1}, \lambda_{2}), & j \in \textrm{Top}_{M_i} \\
	 0 , & j \not\in \textrm{Top}_{M_i},
	\end{cases}
      \end{equation}
      for each fixed $\lambda_1$ and $\lambda_2$.
\textcolor{black}{The biomarkers with non-zero coefficients in $\widehat{\bbeta}_{1}=\widehat{\bbeta}_{1}^{(M_1)}$
  (resp. $\widehat{\bbeta}_{2}^{(M_2)}-\widehat{\bbeta}_{1}^{(M_1)}$)
  are considered as prognostic (resp. predictive) biomarkers, where the choice of $M_1$ and $M_2$ is explained in Section \ref{K1K2M}.} 

 To illustrate the benefits of using an additional thresholding step, we used the dataset described in Section \ref{subsec:gamma_tilde}. Moreover, to simplify the graphical illustrations, we also focus on the case where $\lambda_{1}=\lambda_{2}=\lambda$. 
Figure \ref{fig:beta_filter} \textcolor{black}{in the Supplementary material} displays the number of True Positive (TP) and False Positive (FP) in prognostic and predictive biomarker identification
  with and without the second thresholding. We can see from this figure that the thresholding stage limits the number of false positives.
\textcolor{black}{Note that $\alpha_1$ and $\alpha_2$ are estimated by $\widehat{\widetilde{\alpha}}_{1}$ and
      $\widehat{\widetilde{\alpha}}_{2}$ defined in (\ref{eq:gamma0}).}

\subsection{Choice of the parameters $K_1, K_2$, $M_1$ and $M_2$} \label{K1K2M}
\begin{figure}[!h]
\centering
\includegraphics[scale=0.5]{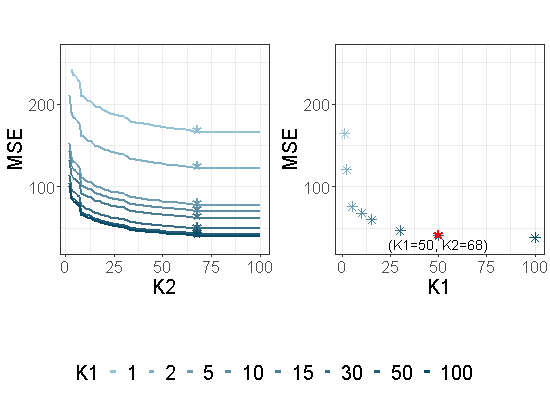}
  \caption{Illustration of how to choose $K_1$ and $K_2$ ($\delta=0.95$), final choice is marked with '\textcolor{red}{*}'. \label{fig:K1K2}}
\end{figure}

For each $(\lambda_1, \lambda_2)$ and each $K_1$, we computed:
\begin{equation}\label{eq:MSEk1k2}
\widetilde{\textrm{MSE}}_{K_1, K_2}(\lambda_1, \lambda_2)=\|\by-\widetilde{\bX}\widehat{\widetilde{\bgamma}}^{(K1, K2)}(\lambda_1, \lambda_2)\|_2^2,
\end{equation}
where $\widehat{\widetilde{\bgamma}}^{(K1, K2)}(\lambda_1, \lambda_2)=(\widehat{\widetilde{\alpha}}_1,\widehat{\widetilde{\alpha}}_2,
\widehat{\widetilde{\bbeta}}_1^{(K_1)'},\widehat{\widetilde{\bbeta}}_2^{(K_2)'})$ defined in (\ref{eq:gamma0}) and
in (\ref{eq:beta_tilde_thresh}). It is displayed in the left part of Figure \ref{fig:K1K2}. 

For each $\lambda_{1}$, $\lambda_{2}$ and a given $\delta\in (0,1)$, the parameter $\widehat{K_2}$ is then chosen as follows for each $K_1$:
\begin{equation*}
\widehat{K_2}(\lambda_{1}, \lambda_{2})=\argmin\left\{K_2\geq 1 \textrm{ s.t. }\frac{\widetilde{\textrm{MSE}}_{(K_1, K_2+1})(\lambda_{1}, \lambda_{2})}{\widetilde{\textrm{MSE}}_{(K_1,K_2)}(\lambda_{1}, \lambda_{2})} \geq\delta\right\}.
\end{equation*}
The $\widehat{K}_2$ associated to each $K_1$ are displayed with '*' in the left part of Figure \ref{fig:K1K2}. Then $\widehat{K_1}$ is chosen by using a similar
criterion:
\begin{equation*}
\widehat{K_1}(\lambda_{1}, \lambda_{2})=\argmin\left\{K_1\geq 1 \textrm{ s.t. }\frac{\widetilde{\textrm{MSE}}_{(K_1+1, \widehat{K}_2})(\lambda_{1}, \lambda_{2})}{\widetilde{\textrm{MSE}}_{(K_1,\widehat{K}_2)}(\lambda_{1}, \lambda_{2})} \geq\delta\right\}.
\end{equation*}
The values of $\widetilde{\textrm{MSE}}_{(K_1,\widehat{K}_2)}(\lambda_{1}, \lambda_{2})$ are displayed in the right part of Figure \ref{fig:K1K2}
in the particular case where $\lambda_{1}=\lambda_{2}=\lambda$, $\delta=0.95$ and with the same dataset as the one used in Section
\ref{subsec:gamma_tilde}. $\widehat{K}_1$ is displayed with a red star. \textcolor{black}{This value of $\delta$ will be used in the following sections. However, choosing $\delta$ in the range (0.9,0.99) does
not have a strong impact on the variable selection performance of our approach.}

\textcolor{black}{ The parameters $\widehat{M}_1$ and $\widehat{M}_2$ are chosen in a similar way except that $\widetilde{\textrm{MSE}}_{K_1, K_2}(\lambda_1, \lambda_2)$
  is replaced by $\widehat{\textrm{MSE}}_{M_1,M_2}(\lambda_1, \lambda_2)$ where:
  $$
  \widehat{\textrm{MSE}}_{M_1,M_2}(\lambda_1, \lambda_2)=\|\by-\bX\widehat{\bgamma}^{(M_1,M_2)}(\lambda_1, \lambda_2)\|_2^2,
  $$
with $\widehat{\bgamma}^{(M_1,M_2)}(\lambda_1, \lambda_2)=(\widehat{\widetilde{\alpha}}_1,\widehat{\widetilde{\alpha}}_2,
\widehat{\bbeta}_1^{(M_1)'},\widehat{\bbeta}_2^{(M_2)'})$ defined in (\ref{eq:gamma0}) and (\ref{eq:beta_hat1}).
In the following, $\widehat{\bgamma}(\lambda_1, \lambda_2)=\widehat{\bgamma}^{(\widehat{M}_1,\widehat{M}_2)}(\lambda_1, \lambda_2)$.
}

\subsection{Estimation of $\bSigma_1$ and $\bSigma_2$}\label{Est_Sigma}

As the empirical correlation matrix is known to be a non accurate estimator of $\bSigma$ when $p$ is larger than $n$, a new estimator has to be used.
Thus, for estimating $\bSigma$ we adopted a cross-validation based method designed by \cite{Boileau2021} and implemented in the \texttt{cvCovEst} R package \citep{Boileau2021R}. This method chooses the estimator having the smallest estimation error among several compared methods \textcolor{black}{(sample correlation matrix,  POET (\cite{POET}) and Tapering (\cite{Cai2010}) as examples)}. Since the samples in treatments $t_1$ and $t_2$ are assumed to be collected from the same population, $\bSigma_1$ and $\bSigma_2$ are assumed to be equal.
%

\subsection{Choice of the parameters $\lambda_1$ and $\lambda_2$} \label{bic}

\begin{figure}[H]
\centering
\includegraphics[scale=0.5]{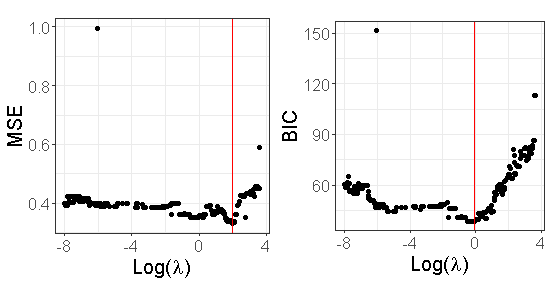}
  \caption{MSE and BIC for all $\lambda$. The $\lambda$ minimizing each criterion is displayed with a vertical line.\label{fig:BIC}}
\end{figure}

For the sake of simplicity, we limit ourselves to the case where $\lambda_1=\lambda_2=\lambda$.
  For choosing $\lambda$ we used BIC (Bayesian Information Criterion) which is widely used in the variable selection field and which
  consists in minimizing the following criterion with respect to $\lambda$:
  $$
  \textrm{BIC}(\lambda)= n\log(\textrm{MSE}(\lambda)/n)+ k(\lambda)\log(n),
  $$
  where $n$ is the total number of samples,
  $
  \textrm{MSE}(\lambda)=\|\by-\bX\widehat{\bgamma}(\lambda)\|_2^2
  $ and $k(\lambda)$ is the number of non null coefficients in the OLS estimator $\widehat{\bgamma}$ obtained by re-estimating only the non null components of $\widehat{\bbeta}_1$ and $\widehat{\bbeta}_2-\widehat{\bbeta}_1$.
  The values of the BIC criterion as well as those of the MSE \textcolor{black}{obtained from the dataset described in Section \ref{subsec:gamma_tilde}}
  are displayed in Figure \ref{fig:BIC}.

Table \ref{tab:BIC} \textcolor{black}{in the supplementary material} provides the True Positive Rate (TPR) and False Positive Rate (FPR) when $\lambda$ is chosen either by minimizing the
MSE or the BIC criterion for this dataset. We can see from this table that both of them have TPR=1 (all true positives are identified). However, the FPR based on the BIC criterion is smaller than the one
obtained by using the MSE.


\section{Numerical experiments}\label{sec:numexp}
\textcolor{black}{This section presents a comprehensive numerical study by comparing the performance of our method with other regularized approaches in terms of prognostic and predictive biomarker selection. Besides the Lasso, we also compared with Elastic Net and Adaptive Lasso since they also take into account the correlations. For Lasso, Elastic Net and Adaptive Lasso,} in order to directly estimate
prognostic and predictive effects, $\bX$ and $\bgamma$ in Model (\ref{lm_simp}) were replaced by
 $$
 \bX^{*}=\begin{bmatrix}
    \bone_{n_1,1} & \bzero_{n_1,1} & \bX_{1} & \bzero_{n_1,p} \\
    \bzero_{n_2,1} & \bone_{n_2,1} &  \bX_{2} & \bX_{2} 
  \end{bmatrix},
  $$
  and $\bgamma^*=(\alpha_1,\alpha_2,\bbeta_1^*,\bbeta_2^*)$, respectively,
 where $\bX_1$ and $\bX_2$ are defined in (\ref{eq:X1_X2}), $\bzero_{i,j}$ (resp. $\bone_{i,j}$) denotes a matrix having $i$ rows and $j$ columns and containing only zeros (resp. ones).
 Note that this is the modeling proposed by \cite{Lipkovich2017}.
\textcolor{black}{The sparsity enforcing constraint was put on the coefficients $\bbeta_1^*$ and $\bbeta_2^*$ which boils down to putting
  a sparsity enforcing constraint on  $\bbeta_1$ and $\bbeta_2-\bbeta_1$.}
 
\subsection{Simulation setting}\label{setting}
All simulated datasets were generated from Model (\ref{lm_simp}) where the $n_1$ ($n_2$) rows of $\bX_{1}$ ($\bX_{2}$) are assumed to be independent Gaussian random vectors with a covariance matrix \textcolor{black}{$\bSigma_{1}=\bSigma_{2}=\bSigma_{bm}$}, and $\bepsilon$ is a standard Gaussian random vector independent of $\bX_{1}$ and $\bX_{2}$. We defined $\bSigma_{bm}$ as: 
\begin{equation}
	     \label{eq:SPAC}
	     \bSigma_{bm}=
	       \begin{bmatrix}
	         \bSigma_{11} &  \bSigma_{12} \\
	         \bSigma_{12}^{T} &  \bSigma_{22}
	       \end{bmatrix}
\end{equation} 
where $\bSigma_{11}$ \textcolor{black}{(resp. $\bSigma_{22}$)} are the correlation matrix of prognostic \textcolor{black}{(resp. non-prognostic)
biomarkers} with off-diagonal entries equal to $a_1$ \textcolor{black}{(resp. $a_3$)}.
  Morever, $\bSigma_{12}$ is the correlation matrix between prognostic and non-prognostic variables with entries equal to $a_2$. In our simulations $(a_1,a_2, a_3)=(0.3, 0.5, 0.7)$, \textcolor{black}{which is a framework proposed by \cite{Xue2017SPAC}.} We checked that the \textcolor{black}{Irrepresentable Condition (IC) of \cite{Zhao:2006} is violated and thus the standard
Lasso cannot recover the positions of the null and non null variables.}
For each dataset we assumed \textcolor{black}{randomized} treatment allocation between standard and experimental arm with a 1:1 ratio, \textit{i.e.} $n_1=n_2=50$. We further assume a relative treatment effect of 1 ($\alpha_{1}=0$ and $\alpha_{2}=1$). The number of biomarkers $p$ varies from 200 to 2000. The number of active biomarkers was set to 10 (\textit{i.e.} 5 purely prognostic biomarkers with $\bbeta_{1j}=\bbeta_{2j}=b_1=1 $ $(j=1,...,5)$ and 5 biomarkers both prognostic and predictive with $\bbeta_{1j}=b_1$ and $\bbeta_{2j}=b_2=2 $ $(j=6,...,10)$). 

\subsection{Evaluation criteria}
We considered several evaluation criteria to assess the performance of the methods in selecting the prognostic and predictive biomarkers: the $\textrm{TPR}_{\textrm{prog}}$ as the true positive rate (i.e. rate of active biomarkers selected) and $\textrm{FPR}_{\textrm{prog}}$ the false positive rate (i.e. rate of inactive biomarkers selected) of the selection of prognostic biomarkers, and similarly for predictive biomarkers with $\textrm{TPR}_{\textrm{pred}}$ and $\textrm{FPR}_{\textrm{pred}}$. We further note $\textrm{TPR}_{\textrm{all}}$ and $\textrm{FPR}_{\textrm{all}}$ the criterion of overall selection among all candidate biomarkers regardless their prognostic or predictive effect. The objective of the selection is to maximize the $\textrm{TPR}_{\textrm{all}}$ and minimize the $\textrm{FPR}_{\textrm{all}}$.
All metrics were calculated by \textcolor{black}{averaging the results of 100 replications for each scenario.}

\subsection{Biomarker selection results}
\textcolor{black}{For the proposed method, different results are presented. $\textrm{PPLasso}_{\Sigma}$ (resp. $\textrm{PPLasso}$) corresponds to the results of the method by considering the true (resp. estimated) matrix $\bSigma_{bm}$. For estimating $\bSigma_{bm}$, we used the approach explained in Section \ref{Est_Sigma}. Two choices of parameters are also given: ``optimal" and ``min(bic)''. The former uses as a value for $\lambda$ the one that maximizes $(\textrm{TPR}_{\textrm{all}}-\textrm{FPR}_{\textrm{all}})$ and the latter uses the approach presented in Section \ref{bic}. In order to compare the performance of our approach to the best performance
  that could be reached by Elastic Net, Lasso and Adaptive Lasso, we used for these methods the ``optimal'' parameters namely those maximizing $(\textrm{TPR}_{\textrm{all}}-\textrm{FPR}_{\textrm{all}})$. All these three methods were implemented with the \texttt{glmnet} R package, the best parameter $\alpha$ involved in Elastic Net was chosen in the set $\{0.1, 0.2,\dots,0.9\}$. The choice of ``min(bic)'' is only applied to our method and corresponds to a choice of $\lambda$ that could be used in
  practical situations. For ease of presentation, the abbreviation EN (resp. AdLasso) refers to Elastic Net (resp. Adaptive Lasso) in the following.} 
\begin{figure}[!h]
	  \begin{center}
	    \includegraphics[scale=0.4]{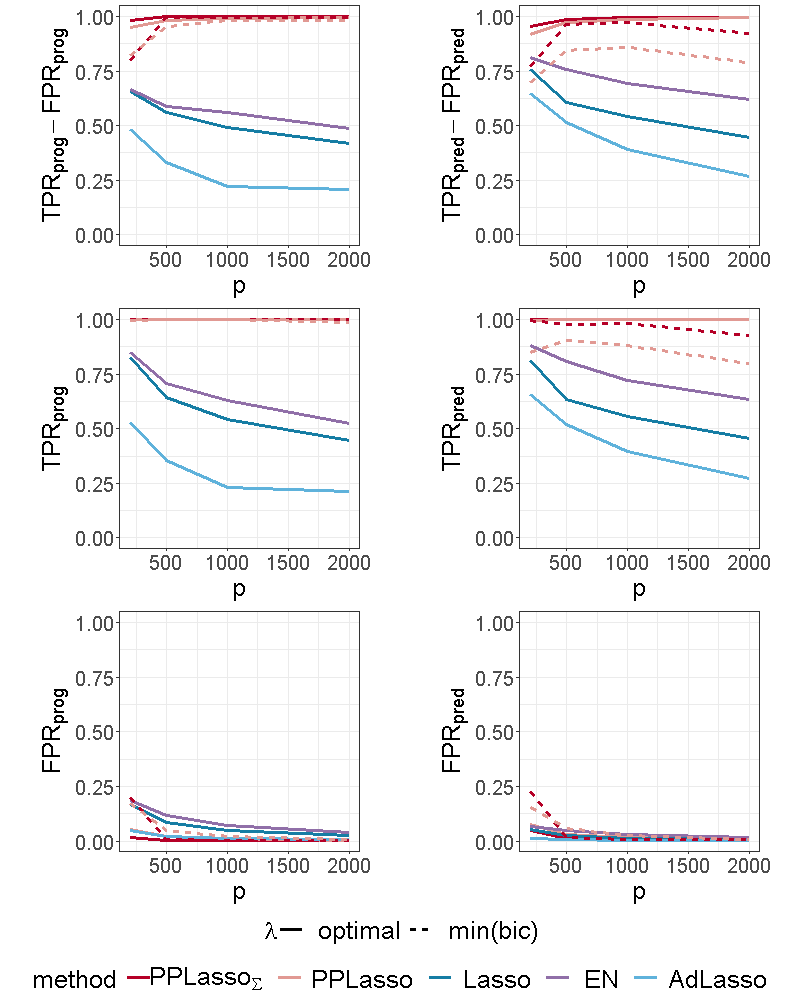}
	  \end{center}
	  \caption{Average of (TPR-FPR) and the corresponding True Positive Rate (TPR) and False Positive Rate (FPR) for prognostic (left) and predictive (right) biomarkers. \label{fig:357_1_1_pred5}}
\end{figure}

\textcolor{black}{Figure} \ref{fig:357_1_1_pred5} shows the selection performance of PPLasso and \textcolor{black}{other compared methods} in the simulation scenario presented in Section \ref{setting}. PPLasso achieved to select all prognostic biomarkers ($\textrm{TPR}_{\textrm{prog}}$ almost 1) even for large $p$, with limited false positive prognostic biomarkers selected. As compared to the optimal $\lambda$ maximizing $(\textrm{TPR}_{\textrm{all}}-\textrm{FPR}_{\textrm{all}})$, the one selected with the BIC tends to select some false positives (average: 33 ($\textrm{FPR}_{\textrm{prog}} = 0.17$) for $p=200$ and 10 ($\textrm{FPR}_{\textrm{prog}} = 0.005$) for $p=2000$). The results
obtained from the oracle and estimated $\bSigma_{bm}$ are comparable. Selection performance of predictive biomarkers is slightly lowered as compared to prognostic biomarkers. \textcolor{black}{Even if the false positive selection is quite similar between prognostic and predictive biomarkers, PPLasso missed some true predictive biomarkers when $\lambda$ is selected with the BIC criterion} (average $\textrm{TPR}_{\textrm{pred}} =$ 0.98 and 0.80 for oracle and estimated $\bSigma_{bm}$, respectively, with $p=2000$). In this scenario where the IC is violated, PPLasso globally \textcolor{black}{outperforms Lasso, Elastic Net and Adaptive Lasso. Although Elastic Net showed higher TPR than Lasso and Adaptive Lasso, they all failed in selecting all truly prognostic and predictive biomarkers, and the number of missed active biomarkers increased with the dimension $p$. For example, for Elastic Net, $\textrm{TPR}_{\textrm{prog}}$ = 0.85 and 0.53, $\textrm{TPR}_{\textrm{pred}}$ = 0.81 and 0.61 for $p = 200$ and $2000$, respectively.}

\subsubsection{Impact of the correlation matrix $\bSigma$}\label{comp1}
To evaluate the impact of the correlation matrix on the selection performance of the methods, additional scenarios are presented where the IC is satisfied:
\begin{enumerate}
\item Compound symmetry structure where all biomarkers are equally \textcolor{black}{correlated  with a correlation $\rho=0.5$};
\item Independent setting where $\bSigma_{bm}$ is the identity matrix. 
\end{enumerate}

\begin{figure}[!h]
	\begin{center}
	    \includegraphics[scale=0.4]{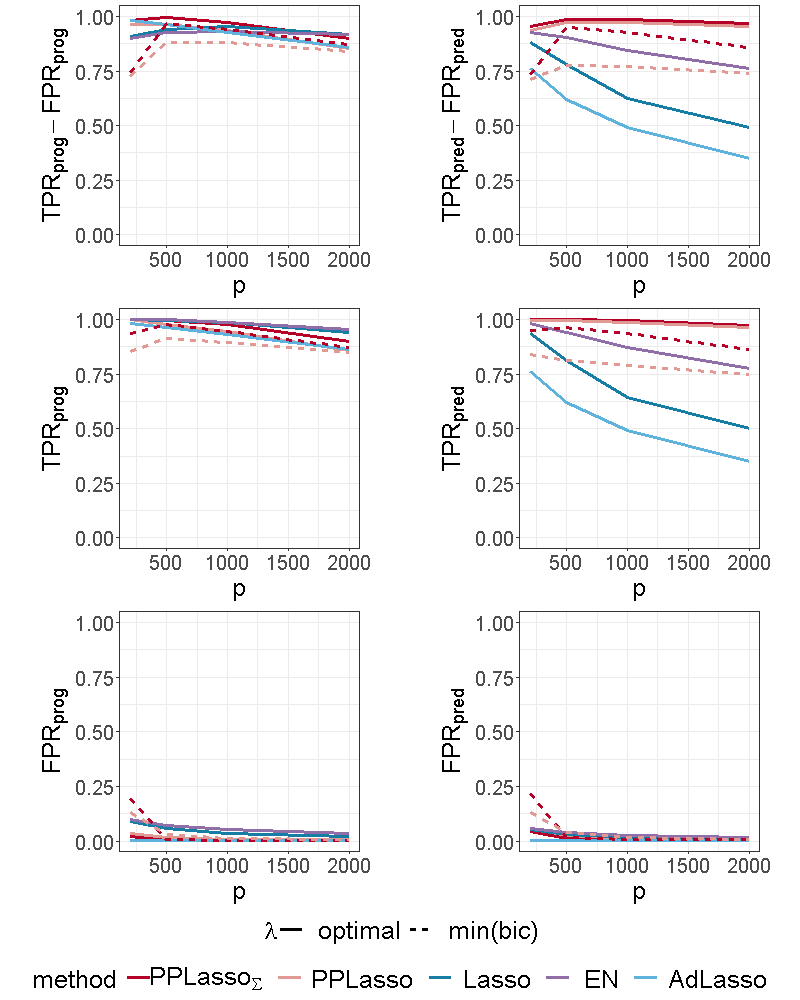}
	\end{center}
	  \caption{Average of (TPR-FPR) and the corresponding True Positive Rate (TPR) and False Positive Rate (FPR) for prognostic (left) and predictive (right) biomarkers
            for the compound symmetry correlation structure.\label{fig:555_1_1_pred5}}
\end{figure}

\textcolor{black}{For the scenario with compound symmetry structure displayed in Figure \ref{fig:555_1_1_pred5}, all the methods successfully identified the true prognostic biomarkers ($\textrm{TPR}_{\textrm{prog}}$ close to 1 even for large $p$) with limited false positive selection.} \textcolor{black}{On the other hand, the compared methods (Lasso, ELastic Net, Adaptive Lasso) missed some predictive biomarkers especially when $p$ increases.}
On the contrary, PPLasso successfully identified almost all predictive biomarkers \textcolor{black}{with the optimal choice of $\lambda$. Moreover, even when $\lambda$ is selected by minimizing the BIC criterion (min(bic)), $\textrm{PPLasso}_{\textrm{est}}$ outperformed Lasso and Adaptive Lasso when $p > 500$ with relatively stable $\textrm{TPR}_{\textrm{pred}}$ and $\textrm{FPR}_\textrm{pred}$ as $p$ increases.}

\begin{figure}[!h]
	  \begin{center}
	    \includegraphics[scale=0.4]{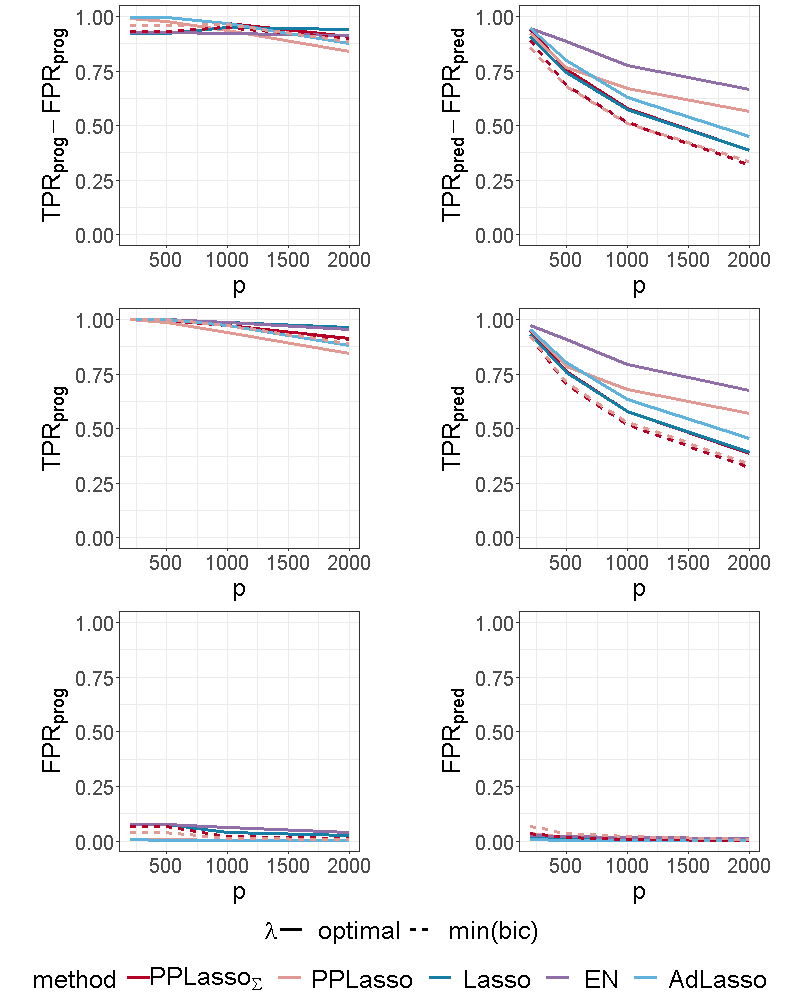}
	  \end{center}
	  \caption{Average of (TPR-FPR) and the corresponding True Positive Rate (TPR) and False Positive Rate (FPR) for prognostic (left) and predictive (right) biomarkers (independent setting). \label{fig:000_1_1_pred5}}
\end{figure}

For the independent setting, as displayed in Figure \ref{fig:000_1_1_pred5}, prognostic biomarkers were globally well identified by all the compared methods with a slightly higher $\textrm{TPR}_{\textrm{prog}}$ for \textcolor{black}{Lasso and ELastic Net} as compared to PPLasso but also with a slightly higher $\textrm{FPR}_{\textrm{prog}}$. With regards to predictive biomarkers, PPLasso using $\bSigma_{bm}$ (oracle) performed also similarly to the Lasso, which is reasonable since no transformation has been used in PPLasso. On the other hand, even if PPLasso with $\lambda$ selected with ``min(bic)'' performed similarly with PPLasso with optimal $\lambda$ for relatively small $p$, the selection performance is altered for large $p$ and \textcolor{black}{even if the performance is higher than Lasso and Adaptive Lasso, it is
  smaller than the one of Elastic Net.} 

 
\subsubsection{Impact of the effect size of active biomarkers}\label{comp2}
To evaluate the impact of the effect size on biomarker selection performance, the scenario presented in Section \ref{setting} was considered with different values of $b_{2}$: 1.5, 2 and 2.5.   

Since the effect size of prognostic biomarkers did not change,
the comparison focused on predictive biomarkers. As expected, the reduction of the effect size makes the biomarker selection harder, \textcolor{black}{especially for Lasso, Elastic Net and Adaptive Lasso where the predictive biomarker selection is limited when $b_{2}=1.5$: for Lasso when $p=2000$, $\textrm{TPR}_{\textrm{pred}}$ = 0.45 (resp. 0.22) for $b_{2}=2$ (resp. 1.5), }see \textcolor{black}{Figure \ref{fig:357_1_1_pred5} and Figure \ref{fig:357_1_0.5_pred5} of the supplementary material.} The selection performance of PPLasso when $\lambda$ is selected with min(bic) is also reduced by decreasing $b_2$, especially when $\bSigma_{bm}$ is also estimated. Nevertheless, the selection performance
\textcolor{black}{of PPLasso} remains better than for the other compared methods \textcolor{black}{for which the performance displayed are associated to the optimal value of $\lambda$.} On the other hand, even with limited effect size, PPLasso with optimal $\lambda$ \textcolor{black}{identified} all predictive biomarkers with very limited false positive selection. When
$b_2$ \textcolor{black}{was increased} to 2.5, the selection performance for all methods is improved and \textcolor{black}{the results for PPLasso with estimated $\lambda$ was close to the ones with the optimal $\lambda$ as displayed in Figure \ref{fig:357_1_1.5_pred5} \textcolor{black}{of the supplementary material}. As compared with PPLasso, for which the selection performance remained stable as $p$ increased, Lasso, Elastic Net and Adaptive Lasso were more impacted by the value of $p$ since the true positive selection decreased as $p$ increased.} \textcolor{black}{As an example, for the Lasso, $\textrm{TPR}_{\textrm{pred}}$ =0.95 (resp. 0.65) for $p = 200$ (resp. 2000).}

\subsubsection{Impact of the number of predictive biomarkers}\label{comp3}
The impact of the number of true predictive biomarkers was assessed by increasing the number of predictive biomarkers from 5 to 10 in the scenario presented in Section \ref{setting}. When the number of predictive biomarkers increased, the impact on PPLasso is almost negligible, especially for prognostic biomarker identification. However, for the other methods, we can see from Figure \ref{fig:357_1_1_pred10} \textcolor{black}{of the supplementary material} that it became even harder to identify predictive biomarkers. $\textrm{TPR}_{\textrm{pred}}$ decreased compared to Figure \ref{fig:357_1_1_pred5}, \textcolor{black}{especially for large $p$ (e.g. $\textrm{TPR}_{\textrm{pred}} = $ 0.12, 0.18, and 0.02 for Lasso, Elastic Net and Adaptive Lasso respectively when $p = 2000$).}

\subsubsection{Impact of the dimension of the dataset}\label{comp4}
\textcolor{black}{In this section, we studied a different sample size: $n$=50 with $n_1=n_2=25$ and a different number of biomarkers: $p$=5000.}

We can see from Figure \ref{fig:357_1_1_pred5_p5000} \textcolor{black}{of the supplementary material} that for $p=5000$, the selection performance of PPLasso is not altered as compared with
$p=2000$ while \textcolor{black}{the compared methods have more difficulties to identify both prognostic and predictive biomarkers.}

  When the sample size is smaller ($n$=50), we can see from Figure \ref{fig:357_1_2_pred5_n50} \textcolor{black}{of the supplementary material} that the ability to identify prognostic and predictive biomarkers decreased for all the methods. However, PPLasso still outperformed \textcolor{black}{the others} with higher $\textrm{TPR}_{\textrm{prog}}$ and $\textrm{TPR}_{\textrm{pred}}$ and lower $\textrm{FPR}_{\textrm{prog}}$ and $\textrm{FPR}_{\textrm{pred}}$.
  

\section{Application to transcriptomic data in RV144 clinical trial}\label{sec:real}
We applied the previously described methods to publicly available transcriptomic data from the RV144 vaccine trial (\cite{Ngarm2009}). This trial showed reduced risk of HIV-1 acquisition by 31.2\% with vaccination with ALVAC and AIDSVAX as compared \textcolor{black}{to} placebo. Transcriptomic profiles of in vitro HIV-1 Env-stimulated peripheral blood mononuclear cells (PBMCs) obtained pre-immunization and 15 days after the immunization (D15) from both 40 vaccinees and 10 placebo recipients were generated to better understand underlying biological mechanisms (\cite{Fourati}, Gene Expression Omnibus accession code: GSE103671).

For illustration purpose, the absolute change at D15 in gene mTOR was considered as \textcolor{black}{the continuous endpoint (response)}. mTOR plays a key role in mTORC1 signaling pathway which has been shown to be associated with risk of HIV-1 acquisition (\cite{Fourati}, \cite{Akbay}). The gene expression has been normalized as in the original publication of \cite{Fourati}. After removing non-annotated genes (LOCxxxx and HS.xxxx), the top 2000 genes with the highest empirical variances were included as candidate biomarkers for prognostic and predictive identification from PPLasso and the compared methods. The penalty parameter $\lambda$ for the Lasso and Adaptive Lasso, the parameters $\lambda$ and $\alpha$ for Elastic Net were selected through the classical cross-validation approach. For PPLasso, $\lambda$ was selected based on the criterion described in Section \ref{bic}.

\begin{figure}[H]
\centering
\includegraphics[scale=0.4]{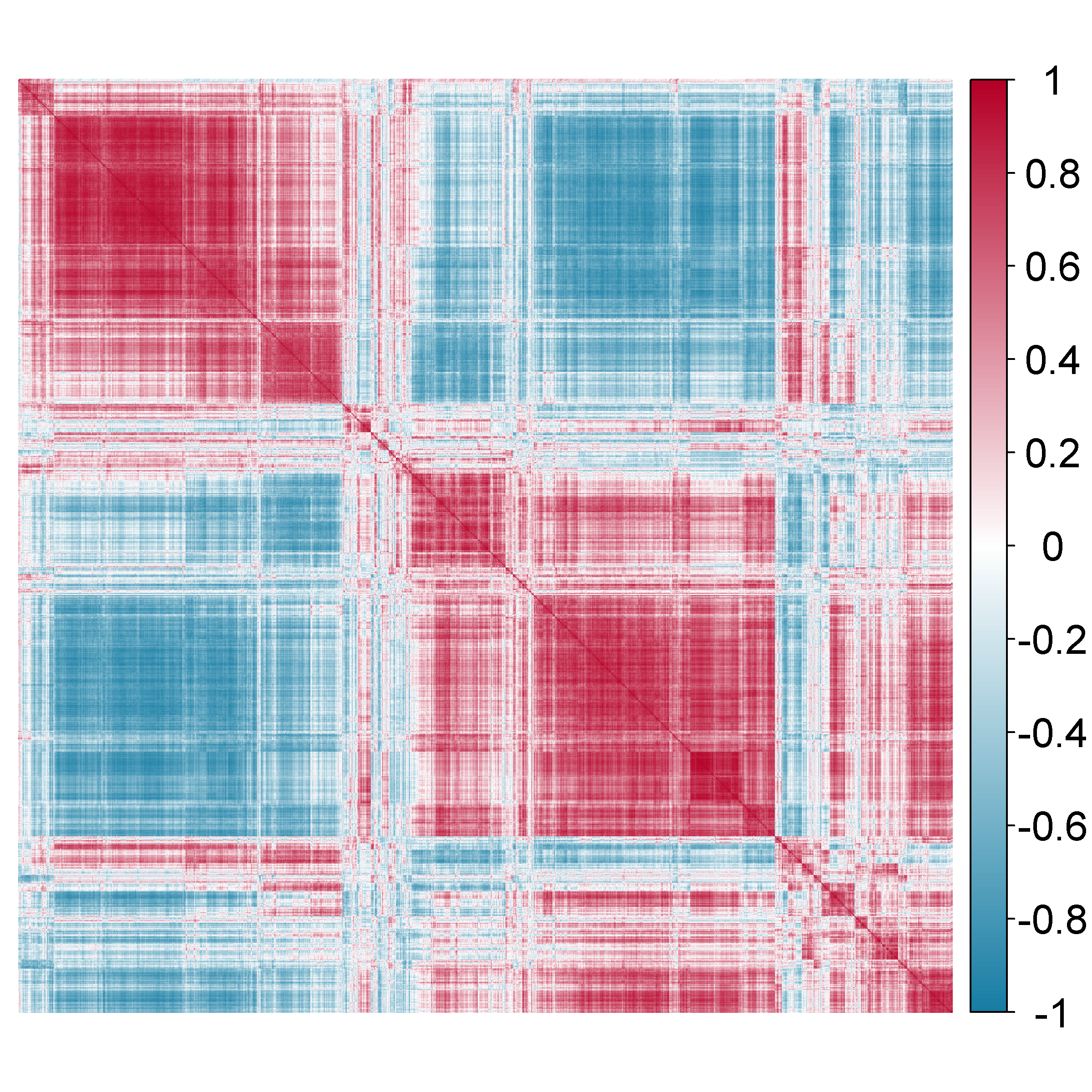}
\caption{Heatmaps of the correlation matrix
  estimated by the \texttt{cvCovEst} R package.
  \label{fig:est_sigma}}
\end{figure}


The estimation of $\bSigma$ was obtained by comparing several candidate estimators from the \texttt{cvCovEst} R package and by selecting the estimator having the smallest estimation error. In this application, the combination of the sample covariance matrix and a dense target matrix (\textit{denseLinearShrinkEst}) derived by \cite{Ledoit2020} provides the smallest estimation error. Figure \ref{fig:est_sigma} displays the estimated $\bSigma$ and highlights the strong correlation between the genes. Table \ref{tab:est_sigma} of the Supplementary material gives details on the compared estimators.\\

\textcolor{black}{Prognostic and predictive genes selected by PPLasso, Lasso, Elastic Net and Adaptive Lasso are listed in Table \ref{Tab:application}. The number of genes selected are similar for all the compared methods, except for a slightly higher number of predictive genes selected by PPLasso. Lasso, Elastic Net and Adaptive Lasso selected very similar sets of prognostic and predictive genes. The intersection between PPLasso and others is moderate (2 prognostic genes (SLAMF7 and TNFRSF6B), 3 predictive genes (YTHDC1, MS4A7 and RPL21)). Interestingly, some genes selected by most methods such as SLAMF7, TNFRSF6B, TNFRSF18 or NUCKS1 have already been discussed in the HIV-1 field. Moreover, among the predictive genes selected by the PPLasso, some are linked to pathways that have been highlighted as possible target for HIV-1 such as BIRC3 and TLR8.}


\begin{table}[]
	  \centering
	    \caption{Selected genes \textcolor{black}{from PPLasso, Lasso, Elastic Net and Adaptive Lasso. Commonly selected genes are in bold.} \label{Tab:application}}
	{\scriptsize{
	    \begin{tabular}{|c|c|c|}
	  \hline
	  & prognostic genes & predictive genes\\\hline
	PPLasso &     \makecell{HAPLN3, \textbf{SLAMF7}, GTF3C5,\\ FAM46A, SH3PXD2B, TM4SF1,\\ \textbf{TNFRSF6B}, TNFRSF18, TRPM2 } &   \makecell{TLR8, \textbf{YTHDC1}, NUCKS1,\\ BIRC3, SLAMF7, NFATC2IP,\\ BOK, MGRN1, KIAA0492, \\SLC25A36, HMGN2, P2RY5,\\ \textbf{RPL21}, \textbf{MS4A7}, RPL12P6 } \\\hline
  Lasso  &\makecell{ DKFZp434K191, NUCKS1, MAFF,\\ \textbf{SLAMF7}, HIST2H2AC, HIST1H4C,\\ IL8, \textbf{TNFRSF6B},\\ TNFRSF18, SCAND1   } &  \makecell{DKFZp434K191, \textbf{YTHDC1},\\ VMO1, BOLA2, HIST1H4C, \\\textbf{RPL21}, \textbf{MS4A7} }   \\\hline
  Elastic Net &     \makecell{DKFZp434K191, NUCKS1,SNURF,\\ MAFF, \textbf{SLAMF7},  IL8,\\ ZBP1, \textbf{TNFRSF6B}, ZAK, \\ TNFRSF18, SCAND1,  NME1-NME2,\\ DNM1L, RNF146, NPEPL1} &   \makecell{DKFZp434K191, \textbf{YTHDC1}, PMP22, \\ VMO1, BOLA2, HIST1H4C, \\\textbf{RPL21}, \textbf{MS4A7},RAB11FIP1} \\\hline
    Adaptive Lasso &     \makecell{ NUCKS1,SNURF, MAFF,\\ \textbf{SLAMF7},  IL8, ZBP1,\\ \textbf{TNFRSF6B}, NME1-NME2,\\ DNM1L, RNF146} &   \makecell{\textbf{YTHDC1}, PMP22, VMO1,\\ BOLA2, HIST1H4C, \textbf{MS4A7},\\\textbf{RPL21}} \\\hline
  
    \end{tabular}
    }}
\end{table}

\section{Conclusion}\label{sec:conclusion}

We propose a new method named PPLasso to simultaneously identify prognostic and predictive biomarkers. PPLasso is particularly interesting for dealing with high dimensional omics data when the biomarkers are highly correlated, which is a framework that has not been thoroughly investigated yet. From various numerical studies with or whithout strong correlation between biomarkers, we highlighted the strength of PPLasso in well identifying both prognostic and predictive biomarkers with limited false positive selection.  The current method is only
dedicated to the analysis of continuous responses through ANCOVA type models. However, it will be the subject of a future work to extend it to other challenging contexts,
such as classification or survival analysis.


\section*{Funding}

This work was supported by the \textcolor{black}{Association Nationale de la Recherche et de la Technologie} (ANRT).

\bibliographystyle{chicago}


\newpage

\section*{Supplementary material}

This supplementary material provides additional numerical experiments, figures and tables for the paper: ``Identification of prognostic and predictive biomarkers in high-dimensional data with PPLasso''.

\begin{figure}
\centering
\includegraphics[scale=0.5]{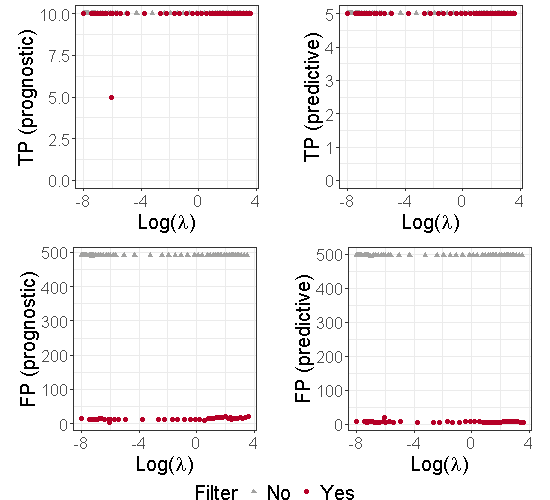}
  \caption{Number of True Positives and True Negatives for $\widehat{\bbeta}$ and $\widehat{\bbeta}_0$ on prognostic/predictive biomarkers.\label{fig:beta_filter}}
\end{figure}

\begin{table}
\centering
\begin{tabular}{l|l|l}
\hline

                & MSE & BIC \\ \hline

TPR(prognostic) &  1.000   &  1.000    \\ 

FPR(prognostic) &  0.038   &  0.024   \\ 

TPR(predictive) &  1.000   &  1.000   \\ 

FPR(predctive)  &  0.008   &  0.006   \\ \hline

\end{tabular}
\newline
 \caption{TPR and FPR associated to prognostic and predictive biomarker identification with the $\lambda$ chosen in Figure \ref{fig:BIC}.
 \label{tab:BIC}}
\end{table}

\begin{figure}[!h]
	  \begin{center}
	    \includegraphics[scale=0.4]{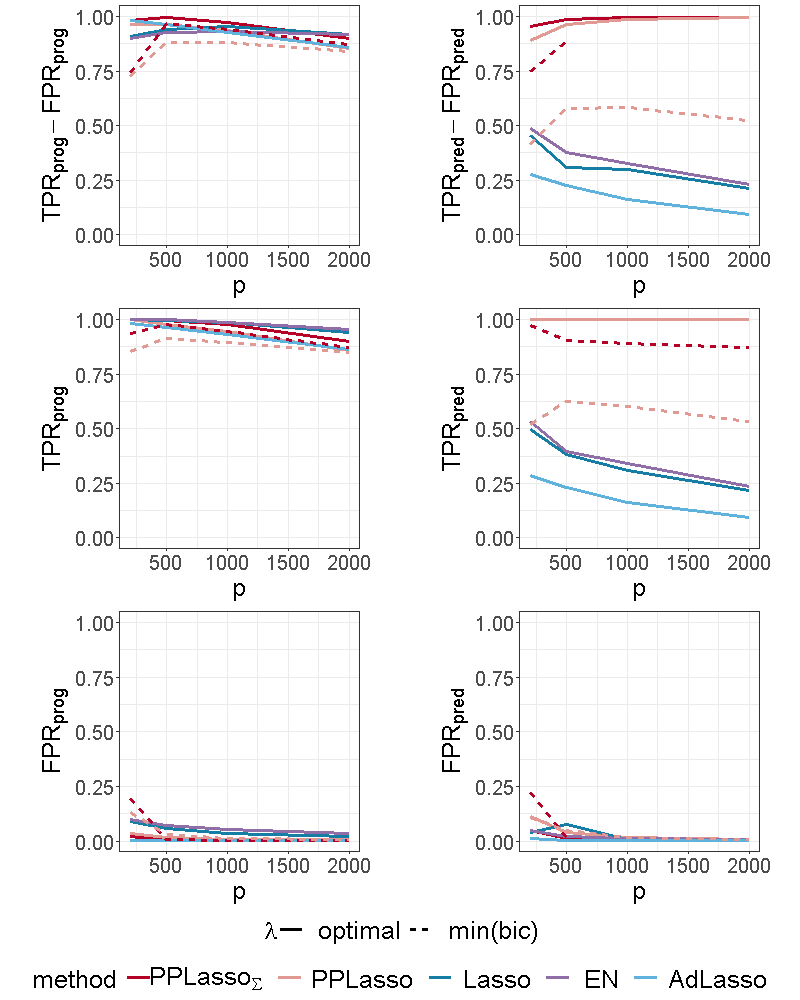}
	  \end{center}
	  \caption{Average of (TPR-FPR) and the corresponding True Positive Rate (TPR) and False Positive Rate (FPR) for prognostic (left) and predictive (right) biomarkers
            ($b_2=1.5$).  \label{fig:357_1_0.5_pred5}}
        \end{figure}
        
\begin{figure}[!h]
	  \begin{center}
	    \includegraphics[scale=0.4]{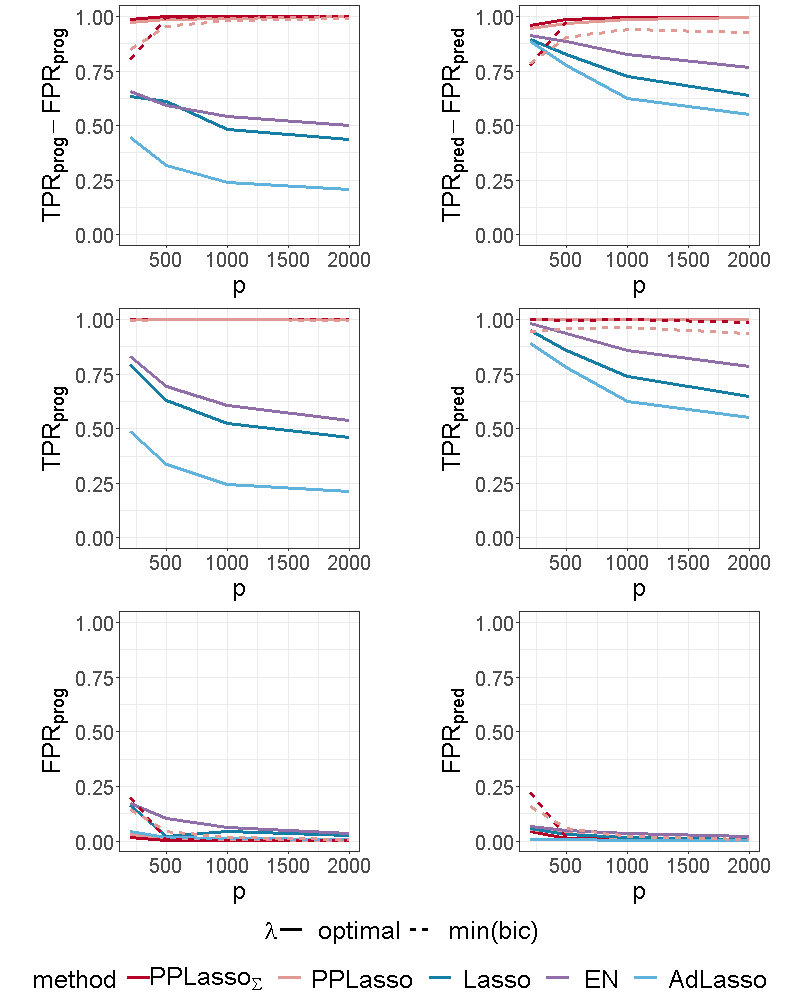}
	  \end{center}
	  \caption{Average of (TPR-FPR) and the corresponding True Positive Rate (TPR) and False Positive Rate (FPR) for prognostic (left) and predictive (right) biomarkers
            ($b_2=2.5$). \label{fig:357_1_1.5_pred5}}
\end{figure}

\begin{figure}[!h]
	  \begin{center}
	    \includegraphics[scale=0.4]{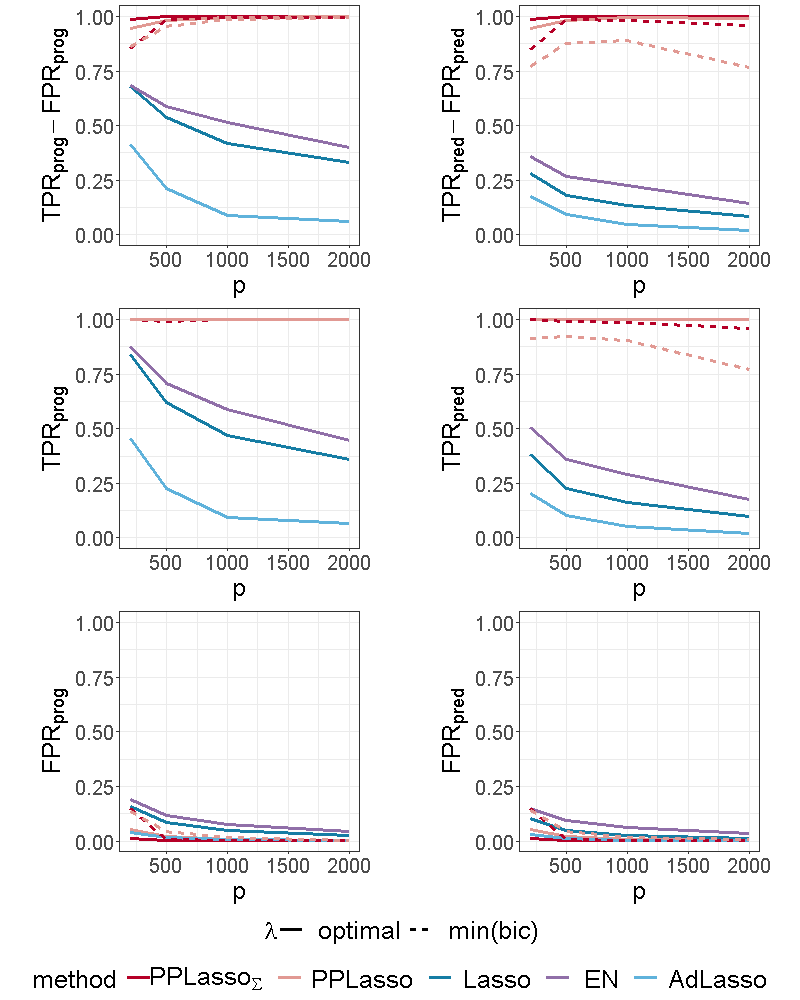}
	  \end{center}
	  \caption{Average of (TPR-FPR) and the corresponding True Positive Rate (TPR) and False Positive Rate (FPR) for prognostic (left) and predictive (right) biomarkers (10 predictive biomarkers). \label{fig:357_1_1_pred10}}
        \end{figure}

        \begin{figure}[!h]
	  \begin{center}
	    \includegraphics[scale=0.4]{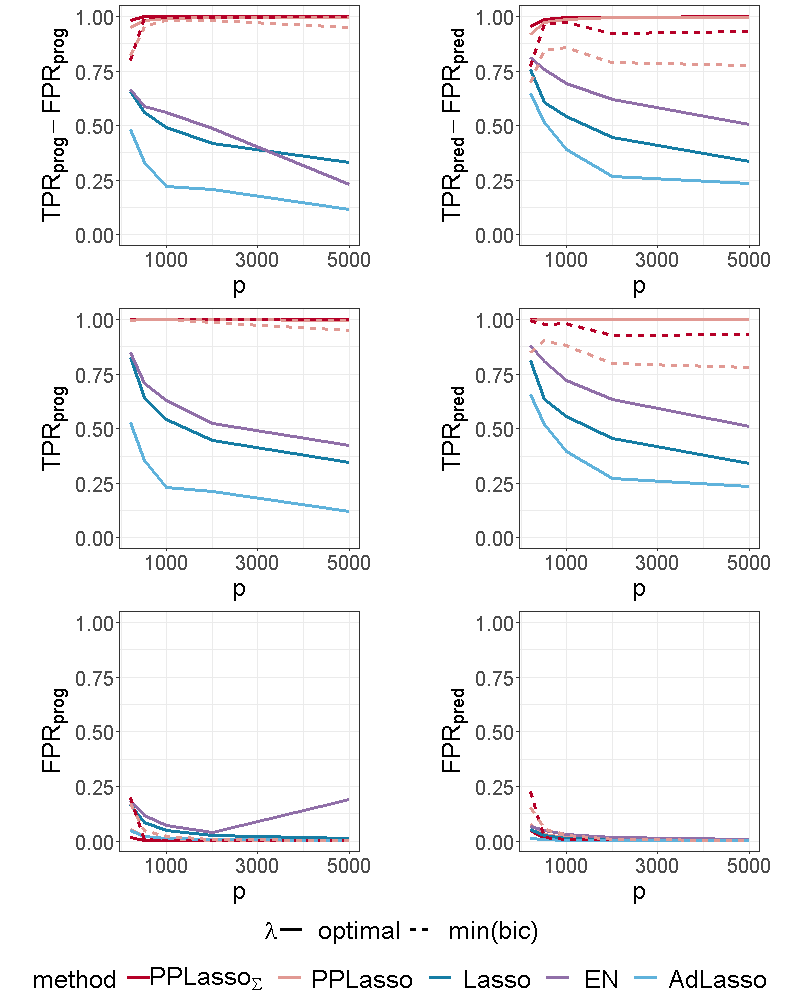}
	  \end{center}
	  \caption{Average of (TPR-FPR) and the corresponding True Positive Rate (TPR) and False Positive Rate (FPR) for prognostic (left) and predictive (right) biomarkers (with $p=5000$).  \label{fig:357_1_1_pred5_p5000}}
\end{figure}
\begin{figure}[!h]
	  \begin{center}
	    \includegraphics[scale=0.4]{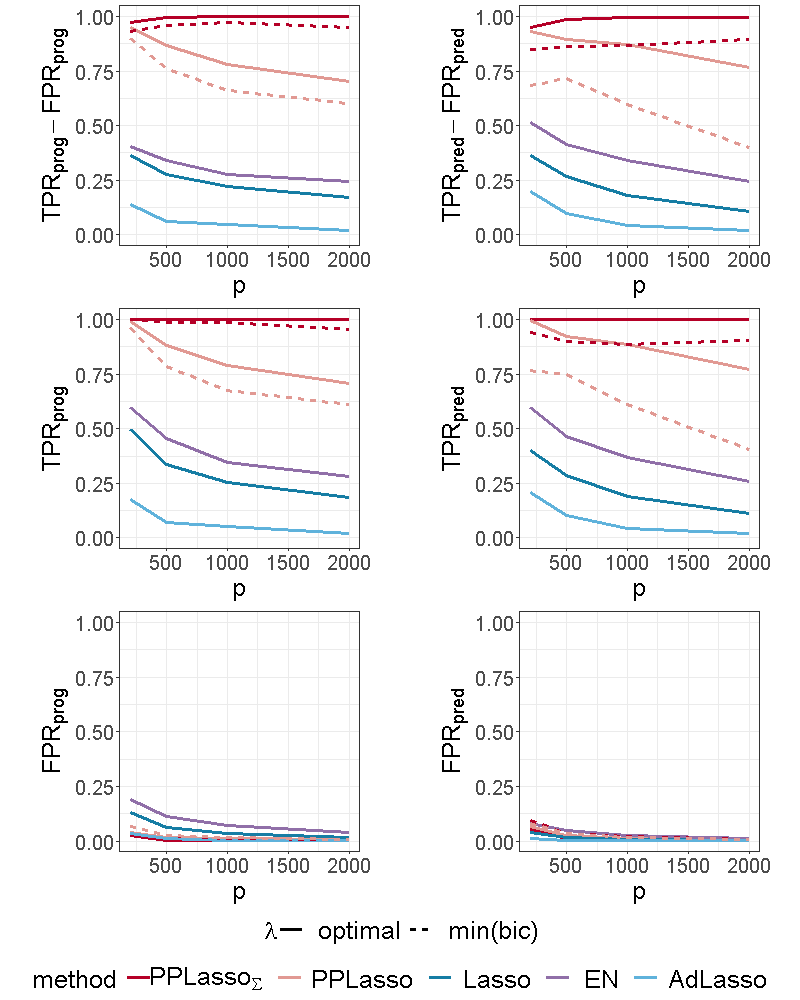}
	  \end{center}
	  \caption{Average of (TPR-FPR) and the corresponding True Positive Rate (TPR) and False Positive Rate (FPR) for prognostic (left) and predictive (right) biomarkers ($n_{1}=n_{2}=25$).  \label{fig:357_1_2_pred5_n50}}
\end{figure}

\begin{table}
\centering
\begin{tabular}{l|c|c}
\hline
\textbf{Estimator}            & \textbf{Hyperparameters}      & \textbf{Empirical risk} \\ \hline
\textit{denseLinearShrinkEst} & \textit{-}              & \textit{102546}            \\ \hline

sampleCovEst         & -              & 102547            \\ \hline

linearShrinkLWEst    & -              & 103496            \\ \hline

poetEst              & lambda=0.1, k=2 & 104522            \\ 

poetEst              & lambda=0.2, k=2 & 105358            \\ 

poetEst              & lambda=0.1, k=1 & 105972            \\ 

poetEst              & lambda=0.2, k=1 & 108222            \\ \hline

thresholdingEst      & gamma=0.2       & 137798            \\ 

thresholdingEst      & gamma=0.4       & 186844            \\ \hline

\end{tabular}
\newline
 \caption{Empirical risk of tested methods with different hyperparameters. 
 \label{tab:est_sigma}}
\end{table}


\end{document}